\documentclass[envcountsame,runningheads]{llncs}
\usepackage{pscproc2}
\usepackage{graphics}
\usepackage{fixltx2e}

\usepackage{xspace,amsmath}
\usepackage{url}

\usepackage{color}
\usepackage{booktabs}

\usepackage{hyperref}

\usepackage{graphics}
\usepackage{epstopdf}
\usepackage{epsfig}

\newcommand{\exclude}[1]{}

\definecolor{red}{rgb}{1.0,0,0}

\DeclareMathOperator{\argmin}{argmin} 

\begin{document}

\title{Faster batched range minimum queries}

\author{Szymon Grabowski, Tomasz Kowalski}
\institute{$^\dag$ Lodz University of Technology, 
           Institute of Applied Computer Science,\\
           Al.\ Politechniki 11, 90--924 {\L}\'od\'z, Poland, \\
           \email{\{sgrabow|tkowals\}@kis.p.lodz.pl}
}

\maketitle

\begin{abstract}
Range Minimum Query (RMQ) is an important building brick of many 
compressed data structures and string matching algorithms.
Although this problem is essentially solved in theory,
with sophisticated data structures allowing for constant time queries,
there are scenarios in which the number of queries, $q$, 
is rather small and given beforehand, 
which encourages to use a simpler approach.
A recent work by Alzamel et al. starts with 
contracting the input array to a much shorter one, 
with its size proportional to $q$.
In this work, we build upon their solution, speeding up 
handling small batches of queries by a factor of 3.8--7.8
(the gap grows with $q$).
The key idea that helped us achieve this advantage is adapting 
the well-known Sparse Table technique to work on blocks, 
with speculative block minima comparisons.
We also propose an even much faster 
(but possibly using more space) 
variant without the array contraction.
\end{abstract}


\section{Introduction}

The Range Minimum Query (RMQ) problem is to preprocess an array so that the position of the minimum element for an arbitrary input interval (specified by a pair of indices) can be acquired efficiently.
More formally, for an array $A[1 \ldots n]$ of objects from a totally ordered universe and two indices $i$ and $j$ such that $1 \leq i \leq j \leq n$, the range minimum query $\mathsf{RMQ_A}(i, j)$ returns $\argmin_{i\leq k\leq j} A[k]$, 
which is the position of a minimum element in $A[i \ldots j]$.
One may alternatively require the position of the leftmost minimum element, i.e., resolve ties in favour of the leftmost such element, but this version of the problem is not widely accepted.
In the following considerations we will assume that $A$ contains integers.

This innocent-looking little problem has quite a rich and vivid history and perhaps even more important applications, in compressed data structures in general, and in text processing in particular.
Solutions for RMQ which are efficient in both query time and preprocessing space and time are building blocks in such succinct data structures as, e.g., suffix trees, two-dimensional grids or ordinal trees. 
They have applications in string mining, document retrieval, bioinformatics, Lempel-Ziv parsing, etc. 
For references to these applications, see~\cite{FH11,FN16}.

The RMQ problem history is related to the LCA (lowest common ancestor) 
problem defined for ordinal trees: 
given nodes $u$ and $v$, return $LCA(u, v)$, which is the lowest node
being an ancestor of both $u$ and $v$. 
Actually, the RMQ problem is linearly equivalent to the LCA problem~\cite{GabowBT84,BFC2000}, by which we mean that both problems 
can be transformed into each other in time linearly proportional 
to the size of the input.
It is relatively easy to notice that 
if the depths of all nodes of tree $T$ visited during an Euler tour 
over the tree are written to array $A$, 
then finding the LCA of nodes $u$ and $v$ is equivalent to finding 
the minimum in the range of $A$ spanned between the first visits to $u$ 
and $v$ during the Euler tour (cf.~\cite[Observation~4]{BFC2000}).
Harel and Tarjan~\cite{HT84} were the first to give $O(n)$-time tree 
preprocessing allowing to answer LCA queries in constant time.
The preprocessing required $O(n)$ words of space.
A significantly simpler algorithm was proposed by Bender and Farach~\cite{BFC2000}, 
with the same time and space complexities.
Further efforts were focused on reducing the space of the LCA/RMQ solution, 
e.g. Sadakane~\cite{Sadakane07} showed that LCAs on a tree of $n$ nodes 
can be handled in constant time using only $2n + o(n)$ bits.
A crowning achievement in this area was the algorithm of 
Fischer and Heun~\cite{FH11}, who showed that RMQs on $A$ can be transformed 
into LCA queries on the succinct tree, and this leads to an RMQ solution 
that also uses $2n + o(n)$ bits and (interestingly) does not access $A$
at query time.

The Fischer and Heun solution, although allowing for constant time RMQ queries, 
is not so efficient in practice: handling one query takes several 
microseconds (see~\cite{FN16}).
Some ingenious algorithmic engineering techniques, 
by Grossi and Ottaviano~\cite{GrossiO13} and by Ferrada and Navarro~\cite{FN16}, 
were proposed to reduce this time, but even the faster of these two~\cite{FN16} 
achieves about $2\mu$s per query\footnote{On an Intel Xeon 2.4\,GHz, 
running on one core (H.~Ferrada, personal comm.).}.

Very recently, Alzamel et al.~\cite{ACIP17} (implicitly) posed 
an interesting question:
why should we use any of these sophisticated data structures for RMQ
when the number of queries is relatively small and building the index 
(even in linear time, but with a large constant) and answering then 
the queries (even in constant time each, but again with a large constant) 
may not amortize?
A separate, but also important point is that if we can replace a heavy tool 
with a simpler substitute (even if of limited applicability), 
new ideas may percolate from academia to software industry.
Of course, if the queries $[\ell_i, r_i]$ are given one by one, 
we cannot answer them faster than in the trivial 
$O(r_i - \ell_i + 1) = O(n)$ time for each, 
but the problem becomes interesting if they are known beforehand.
The scenario is thus offline 
(we can also speak about {\em batched queries} or {\em bulk queries}).
Batched range minima (and batched LCA queries) have applications 
in string mining~\cite{FMV08}, text indexing and various non-standard pattern 
matching problems, for details see~\cite[Section~5]{ACIP17}.

As the ideas from Alzamel et al.~\cite{ACIP17} are a starting point for our 
solution and we directly compete with them, we dedicate the next section 
to presenting them.

We use a standard notation in the paper.
All logarithms are of base 2.
If not stated otherwise, the space usage is expressed in words.

\section{The Alzamel et al. algorithm}

Following~\cite{AS14} (see the proof of Lemma 2), the Alzamel et al. approach 
starts from contracting the array $A$ into $O(q)$ entries.
The key observation is that if no query starts or ends with an index $i$ and $i+1$, 
then, if $A[i] \neq A[i+1]$, $\max(A[i], A[i+1])$ will not be the answer to 
any of the queries from the batch.
This can be generalized into continuous regions of $A$.
Alzamel et al. mark the elements of $A$ which are either a left or a right endpoint
of any query and create a new array $A_Q$: 
for each marked position in $A$ its original value is copied into $A_Q$,
while each maximal block in $A$ that does not contain a marked position is replaced 
by a single entry, its minimum. 
The relative order of the elements copied from $A$ is preserved in $A_Q$, 
that is, in $A_Q$ the marked elements are interweaved with representatives of 
non-marked regions between them.
As each of $q$ queries is a pair of endpoints, $A_Q$ contains up to $4q + 1$ elements 
(repeating endpoint positions imply a smaller size of $A_Q$, but for relative small 
batches of random queries this effect is rather negligible).
In an auxiliary array the function mapping from the indices of $A_Q$ into the original 
positions in $A$ is also kept.

For the contracted data, three procedures are proposed. 
Two of them, one offline and one online, are based on existing RMQ/LCA algorithms 
with linear preprocessing costs and constant time queries. 
Their practical performance is not competitive though.
The more interesting variant, \textsf{ST-RMQ\textsubscript{CON}}, 
achieves $O(n + q\log q)$ time\footnote{Written consistently as $n + O(q\log q)$ 
in the cited work, to stress that the constant associated with scanning the original 
array $A$ is low.}.
The required space (for all variants), on top of the input array $A$ and the list of 
queries $Q$, is claimed to be $O(q)$, but a more careful look into the algorithm 
(and the published code) reveals that in the implementation of the contracting step 
the top bits of the entries of $A$ are used for marking.
There is nothing wrong in such a bit-stealing technique, from a practical 
point\footnote{One of the authors of the current work also practiced it in 
a variant of the SamSAMi full-text index~\cite[Section~2.3]{GR17}.}, 
but those top bits may not always be available and thus in theory the space 
should be expressed as $O(q)$ words plus $O(n)$ bits.

We come back to the \textsf{ST-RMQ\textsubscript{CON}} algorithm.
Bender and Farach~\cite{BFC2000} made a simple observation: 
as the minimum in a range $R$ is the minimum over the minima of arbitrary 
ranges (or subsets) in $R$ with the only requirement that the whole $R$ 
is covered, for an array $A$ of size $n$ it is enough to precompute 
the minima for (only) $O(n\log n)$ ranges to handle any RMQ.
More precisely, for each left endpoint $A[i]$ we compute the minima 
for all valid $A[i \ldots i + 2^k-1]$ ($k = 0, 1, \ldots$) ranges, 
and then for any $A[i \ldots j]$ it is enough to compute the minimum 
of two already computed minima: 
for $A[i \ldots i + 2^{k'} - 1]$ and $A[j - 2^{k'} + 1 \ldots j]$, 
where $k' = \lfloor \log(j - i) \rfloor$.
Applying this technique for the contracted array would yield $O(q\log q)$ 
time and space for this step.
Finally, all the queries can be answered with the described technique, 
in $O(q)$ time.
In the cited work, however, the last two steps are performed together, 
with re-use of the array storing the minima.
Due to this clever trick, the size of the helper array is only $O(q)$.

\section{Our algorithms}
\label{sec:our}

\subsection{Block-based Sparse Table with the input array contraction}
\label{sec:our1}
On a high level, our first algorithm consists of the following four steps:

\begin{enumerate}
\item Sort the queries and remap them with respect 
to the contracted array's indices (to be obtained in step 2).
\item Contract $A$ to obtain $A_Q$ of size $O(q)$ (integers).
\item Divide $A_Q$ into equal blocks of size $k$ and for each block $B_j$ 
(where $j = 1, 2, \ldots$) find and store the positions of $O(\log q)$ minima, 
where $i$th value ($i = 1, 2, \dots $) is the minimum of 
$A_Q[1 + (j-1)k \ldots (j-1)k + (2^{i-1} - k)k]$, 
i.e., the minimum over a span of $2^{i-1}$ blocks, where the leftmost 
block is $B_j$.
\item For each query $[\ell_i, r_i]$, find the minimum $m'_i$ over the 
largest span of blocks fully included in the query and not containing 
the query endpoints.
Then, read the minimum of the block to which $\ell_i$ belongs 
and the minimum of the block to which $r_i$ belongs; 
only if any of them is less than $m'_i$, then scan 
(at most) $O(k)$ cells of $A_Q$ to find the true minimum and return 
its position.
\end{enumerate}

In the following paragraphs we are going to describe those steps in more detail, 
also pointing out the differences between our solution and Alzamel et al.'s one.

{\em (1) Sorting/remapping queries.}
Each of the $2q$ query endpoints is represented as a pair of 32-bit integers: 
its value (position in $A$) and its index in the query list $Q$.
The former 4-byte part is the key for the sort while the latter 4 bytes 
are satellite data.
In the serial implementation, we use kxsort\footnote{\url{https://github.com/voutcn/kxsort}}, 
an efficient MSD radix sort variant. 
In the parallel implementation, our choice was 
Multiway-Mergesort Exact variant implemented in GNU libstdc++ 
parallel mode library\footnote{\url{https://gcc.gnu.org/onlinedocs/libstdc++/manual/parallel_mode.html}}.
As a result, we obtain a sorted endpoint list $E[1 \dots 2q]$, 
where $E_i = (E_i^x, E_i^y)$ and $E_{i+1}^x \geq E_i^x$.
Alzamel et al. do not sort the queries, which is however possible 
due to marking bits in $A$.

{\em (2) Creating $A_Q$.}
Our contracted array $A_Q$ contains the minima of all areas 
$A[E_i^x \ldots E_{i+1}^x]$, in order of growing $i$.
$A_Q$ in our implementation contains thus (up to) $2q-1$ entries, twice less 
than in Alzamel et al.'s solution.
Like in the preceding solution, we also keep a helper array 
mapping from the indices of $A_Q$ into the original positions in $A$.

{\em (3) Sparse Table on blocks.}
Here we basically follow Alzamel et al. in their \textsf{ST-RMQ\textsubscript{CON}} variant, 
with the only difference that we work on blocks rather than individual elements of $A_Q$.
For this reason, this step takes 
$O(q + (q/k)\log(q/k)) = O(q(1 + \log(q/k) / k))$ time and $O((q/k)\log(q/k))$ 
space.
The default value of $k$, used in the experiments, is 512.

{\em (4) Answering queries.}
Clearly, the smaller of two accessed minima in the Sparse Table technique 
is the minimum over the largest span of blocks fully included in the query 
and not containing the query endpoints.
To find the minimum over the whole query we perform {\em speculative reads} 
of the two minima of the extreme blocks of our query.
Only if at least one of those values is smaller than the current minimum, 
we need to scan a block (or both blocks) in $O(k)$ time.
This case is however rare for an appropriate value of $k$.
This simple idea is crucial for the overall performance of our scheme.
In the worst case, we spend $O(k)$ per query here, 
yet on average, assuming uniformly random queries over $A$, 
the time is $O((k/q) \times k + (1-k/q) \times 1) = O(1 + k^2/q)$, 
which is $O(1)$ for $k = O(\sqrt{q})$.

Let us sum up the time (for a serial implementation) and space costs.
A scan over array $A$ is performed once, in $O(n)$ time.
The radix sort applied to our data of $2q$ integers from $\{1, \ldots, n\}$
takes (in theory) $O(q \max(\log n/\log q, 1))$ time.
Alternatively, introsort from C++ standard library (i.e., 
the std::sort function) would yield $O(q\log q)$ time.
To simplify notation, the $Sort(q)$ term will further be used 
to denote the time to sort the queries 
and we also introduce $q' = q/k$.
$A_Q$ is created in $O(q)$ time.
Building the Sparse Table on blocks adds $O(q + q'\log q')$ time.
Finally, answering queries requires $O(qk)$ time in the worst case 
and $O(q + k^2)$ time on average.
In total, we have $O(n + Sort(q) + q'\log q' + qk)$ time in the worst case.
The extra space is $O(q'\log q')$.

Let us now consider a generalization of the doubling technique in Sparse Table 
(a variant that we have not implemented).
Instead of using powers of 2 in the 
formula $A_Q[1 + (j-1)k \ldots (j-1)k + (2^{i-1} - k)k]$, 
we use powers of an arbitrary integer $\ell \geq 2$
(in a real implementation it is convenient to assume that $\ell$ 
is a power of 2, e.g., $\ell = 16$).
Then, the minimum over a range will be calculated as a minimum 
over $\ell$ precomputed values.
Overall we obtain 
$O(n + Sort(q) + q'\log q'/\log\ell + q\ell + qk)$ worst-case time, 
which is minimized for 
$\ell = \max(\log q' / (k\log\log q'), 2)$.
With $k$ small enough to have $\ell = \log q' / (k\log\log q')$, 
we obtain
$O(n + Sort(q) + q'\log q' / \log\log q' + qk)$ overall time 
and the required extra space is $O(q'\log q' / \log\log q')$.

If we focus on the average case, where the last additive term 
of the worst-case time turns into $k^2/q$, 
it is best to take $k = \sqrt{q}$, which implies $\ell = 2$.
In other words, this idea has its niche only considering 
the worst-case time, 
where for a small enough $k$ both the time and the space 
of the standard block-based Sparse Table solution are improved.

\subsection{Block-based Sparse Table with no input array contraction}
\label{sec:our2}

This algorithm greatly simplifies the one from the previous subsection:
we do not contract the array $A$ and thus also have no need to sort the queries.
Basically, we reduce the previous variant to the last two stages.
Naturally, this comes at a price: the extra space usage becomes 
$O((n/k)\log(n/k))$ 
(yet the optimal choice of $k$ may be different, closer to $\sqrt{n}$).
Experiments will show that such a simple idea offers very competitive RMQ times.

Let us focus on the space and time complexities for this variant, 
for both the worst and the average case.
The analysis resembles the one from the previous subsection.
We have two parameters, $n$ and $k$, and two stages of the algorithm.
The former stage takes $O(n + (n/k)\log(n/k))$ time, 
the latter takes $O(qk)$ time in the worst case and 
$O(q(1 + k^2/n))$ on average (which is $O(q)$ if $k = O(\sqrt{n})$).
In total we have $O(n + (n/k)\log(n/k) + qk)$ time in the worst case
and $O(n + (n/k)\log(n/k) + q)$ time on average, 
provided in the latter case that $k = O(\sqrt{n})$.
The space 
is $O((n/k)\log(n/k))$.
To minimize both the time and the space for the average case 
we set $k = \Theta(\sqrt{n})$.
Then the average time becomes $O(n + \sqrt{n}\log\sqrt{n} + q) = O(n + q)$ 
and the space is $O(\sqrt{n}\log n)$.

\subsection{Multi-level block-based Sparse Table}
\label{sec:our3}

The variant from Subsection~\ref{sec:our2} can be generalized to 
multiple block levels.
We start from the simplest case, replacing one level of blocks with two 
levels.

The idea is to compute minima for $n/k_1$ non-overlapping blocks of size $k_1$ 
and then apply the doubling technique from Sparse Table on larger blocks, 
of size $k_2$.
We assume that $k_1$ divides $k_2$.

The first stage, finding the minima for blocks of size $k_1$, takes $O(n)$ time.
The second stage, working on blocks of size $k_2$, 
takes $O(n/k_1 + (n/k_2)\log(n/k_2))$ time.
The third stage answers the queries; if we are unlucky and 
one or two blocks of size $k_2$ have to be scanned, the procedure is sped up 
with aid of the precomputed minima for the blocks of size $k_1$.
The query answering takes thus $O(q(k_2/k_1 + k_1))$ time in the worst case 
and $O(q)$ time on average if $(k_2 / n) \times (k_2/k_1 + k_1) = O(1)$.
The condition on the average case becomes clear when we notice that 
the probability of the unlucky case is $\Theta(k_2/n)$ 
and checking (up to) two blocks takes $O(k_2 / k_1 + k_1)$ time.
Fulfilling the given condition implies that 
$k_1 k_2 = O(n)$ and $k_2 / k_1 = O(n/k_2)$.

Our goal is to find such $k_1$ and $k_2$ that the extra space is minimized 
but the average time of $O(n + q)$ preserved.
To this end, we set $k_1 = \sqrt{n} / \log^{1/3}n$, $k_2 = \sqrt{n} \log^{2/3}n$, 
and for these values the average time becomes 
$O(n + n/k_1 + (n/k_2)\log(n/k_2) + q) =
O(n + q)$.
The space is $O(n/k_1 + (n/k_2)\log(n/k_2)) = 
O(\sqrt{n}\log^{1/3}n)$.

Note that we preserved the average time of the variant 
from Subsection~\ref{sec:our2} and reduced the extra space 
by a factor of $\log^{2/3}n$.
Note also that the space complexity cannot be reduced for any 
other pair of $k_1$ and $k_2$ such that
$k_1 k_2 = O(n)$.

We can generalize the presented scheme to have $h \geq 2$ levels.
To this end, we choose $h$ parameters, $k_1 < \ldots < k_h$, 
such that each $k_i$ divides $k_{i+1}$.
The minima for non-overlapping blocks of size $k_i$, $1 \leq i < h$, 
are first computed, 
and then also the minima for blocks of size $k_h$, 
their doubles, quadruples, and so on.
The $O(q)$ average time for query answering now requires that 
$(k_h / n) \times (k_h / k_{h-1} + k_{h-1} / k_{h-2} + \ldots + k_2 / k_1 + k_1) = O(1)$.
We set $k_1 = \sqrt{n} / \log^{1/(h+1)}n$ 
and $k_i = \sqrt{n} \log^{(i-1)/(h-1) - 1/(h+1)}n$ for all $2 \leq i \leq h$, 
which gives $k_h = \sqrt{n} \log^{h/(h+1)} n$.
Let us suppose that $h = O(\log\log n)$.
The aforementioned condition is fulfilled, the average time is $O(n + q)$, 
and the space is $O(n/k_1 + n/k_2 + \ldots + n/k_{h-1} + (n/k_h)\log(n/k_h)) = 
O(n/k_1 + (n/k_h)\log(n/k_h)n) = O(\sqrt{n} \log^{1/(h+1)}n)$.
By setting $h = \log\log n - 1$ we obtain $O(\sqrt{n})$ words of space.


\begin{figure}[pt]
\centerline{
\includegraphics[width=0.495\textwidth,scale=1.0]{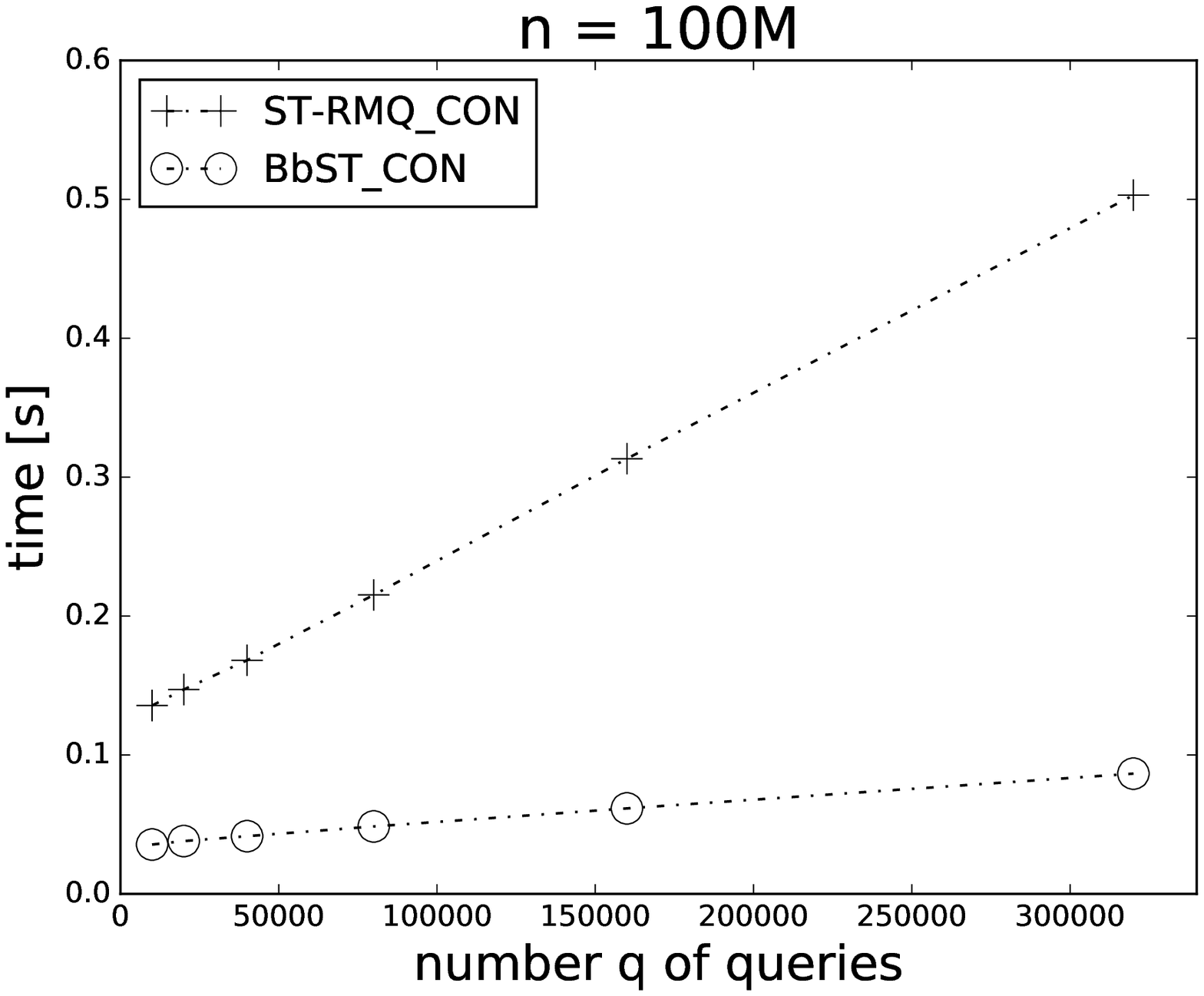}
\includegraphics[width=0.495\textwidth,scale=1.0]{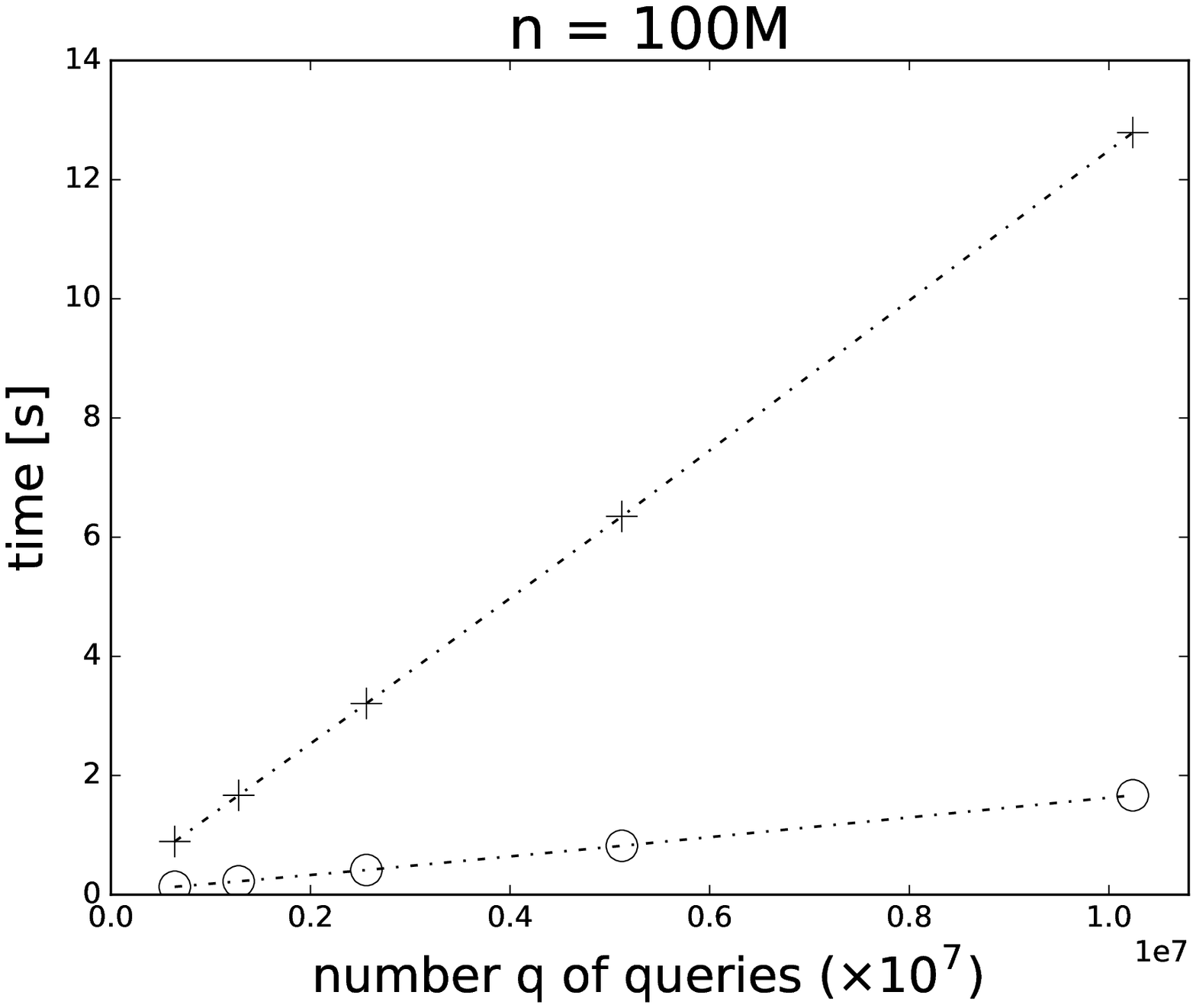}
}
\centerline{
\includegraphics[width=0.495\textwidth,scale=1.0]{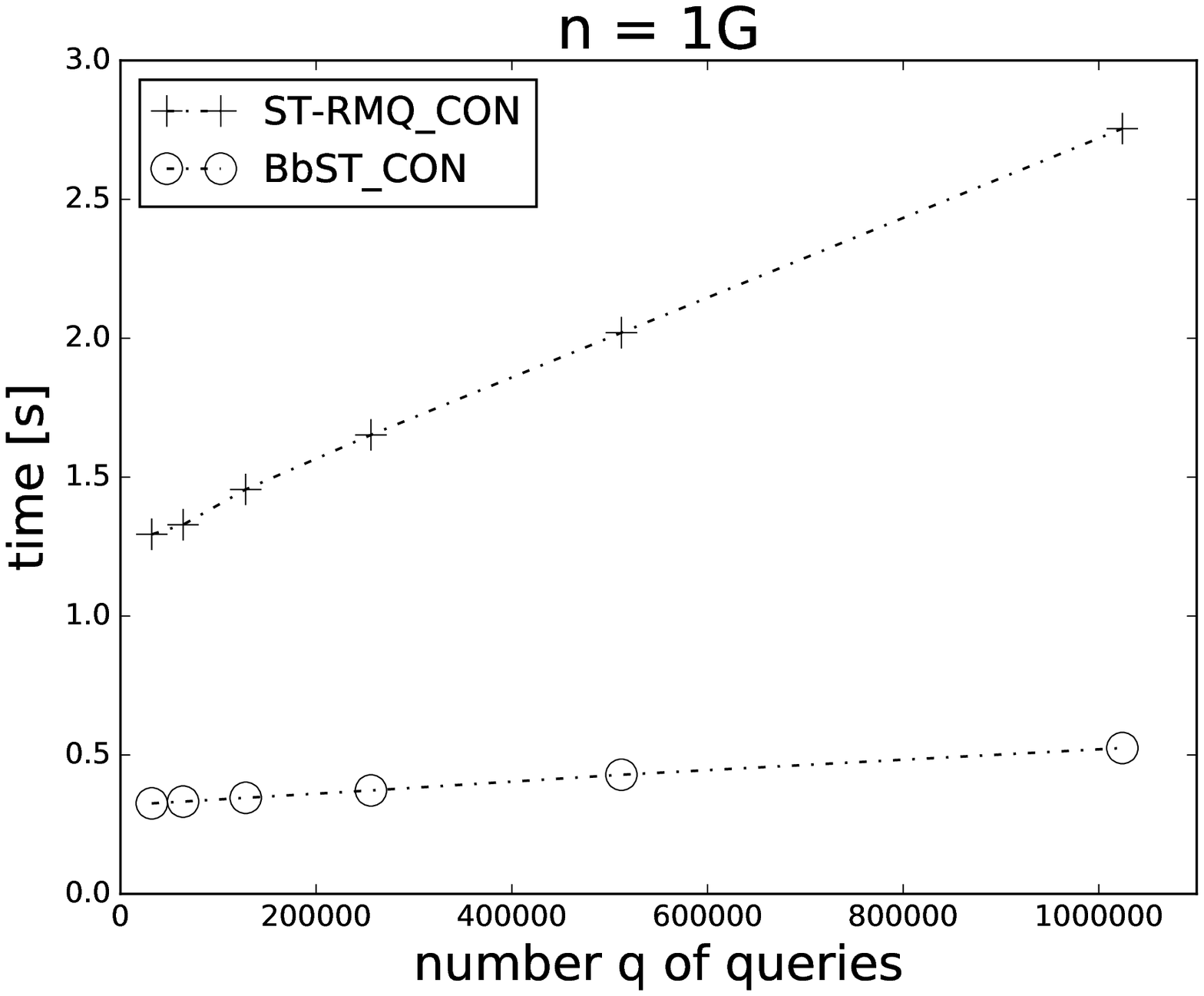}
\includegraphics[width=0.495\textwidth,scale=1.0]{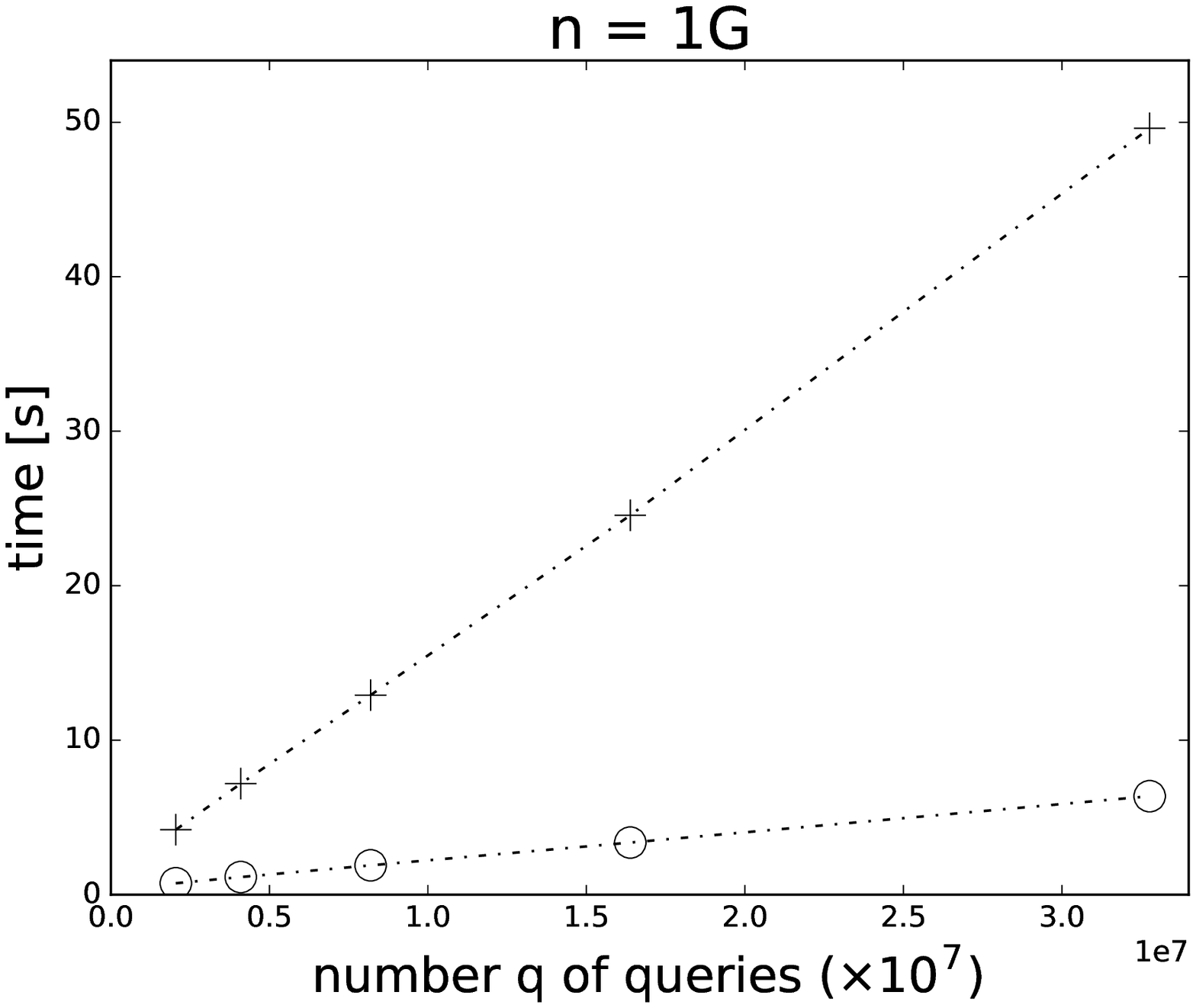}
}
\caption[Fig1]
{Running times for \textsf{ST-RMQ\textsubscript{CON}} and \textsf{BbST\textsubscript{CON}} 
with varying number of queries $q$, 
from $\sqrt{n}$ to $32\sqrt{n}$ (left figures) 
and from $64\sqrt{n}$ to $1024\sqrt{n}$ (right figures), 
where $n$ is 100 million (top figures) or 1 billion (bottom figures)}
\label{fig:main}
\end{figure}

\section{Experimental results}

In the experiments, we followed the methodology from~\cite{ACIP17}.
The array $A$ stores a permutation of $\{1, \ldots, n\}$, 
obtained from the initially increasing sequence 
by swapping $n/2$ randomly selected pairs of elements.
The queries are pairs of the form $(\ell_i, r_i)$, 
where $\ell_i$ and $r_i$ are uniformly randomly drawn from 
$\{1, \ldots, n\}$ 
and if it happens that the former index is greater than the latter, 
they are swapped.
The number of queries $q$ varies from $\sqrt{n}$ to $1024\sqrt{n}$, 
doubling each time (in~\cite{ACIP17} they stop at $q = 128\sqrt{n}$).

Our first algorithm, \textsf{BbST\textsubscript{CON}} (Block based Sparse Table with Contraction), 
was implemented 
in C++ and compiled with 32-bit gcc 6.3.0 with \texttt{-O3 -mavx -fopenmp} 
switches.
Its source codes can be downloaded from \url{https://github.com/kowallus/BbST}.
The experiments were conducted on a desktop PC equipped with 
a 4-core Intel i7 4790 3.6\,GHz CPU and 32\,GB of 1600\,MHz DDR3 RAM (9-9-9-24), 
running Windows 10 Professional.
All presented timings in all tests are medians of 7 runs, 
with cache flushes in between.

In the first experiment we compare \textsf{BbST\textsubscript{CON}} with default settings 
($k = 512$, kxsort in the first stage)
against \textsf{ST-RMQ\textsubscript{CON}} (Fig.~\ref{fig:main}).
Two sizes of the input array $A$ are used, 100 million and 1 billion.
The left figures present the execution times for small values of $q$ 
while the right ones correspond to bigger values of $q$.
We can see that the relative advantage of \textsf{BbST\textsubscript{CON}} over 
\textsf{ST-RMQ\textsubscript{CON}} grows with the number of queries, 
which in part can be attributed to using a fixed value of $k$ 
(the selection was leaned somewhat toward larger values of $q$).
In any case, our algorithm is several times faster than its predecessor.

Table~\ref{table:partial} contains some profiling data.
Namely, cumulative percentages of the execution times for the four successive stages 
(cf.~\ref{sec:our1}) of \textsf{BbST\textsubscript{CON}} with default settings, are shown.
Unsurprisingly, for a growing number of queries the relative impact of 
the sorting stage (labeled as stage 1) grows, otherwise the array contraction (stage 2) 
is dominating.
The last two stages are always of minor importance in these tests.

\begin{table}[t!]
\centering
\begin{tabular}{rrrrr}
\hline
~~$q$ (in 1000s)~~~~~~& stage 1~~~& stages 1--2~~~& stages 1--3~~~& stages 1--4~~~\\
\hline
\multicolumn{5}{c}{$n = 100,000,000$}  \\
\hline
~~~10~~~~~~&  1.4~~~& 95.9~~~& 95.9~~~& 100.0~~~\\
~~320~~~~~~& 23.5~~~& 92.5~~~& 93.0~~~& 100.0~~~\\
10240~~~~~~& 65.8~~~& 88.3~~~& 89.1~~~& 100.0~~~\\
\hline
\multicolumn{5}{c}{$n = 1,000,000,000$}  \\
\hline
~~~32~~~~~~&  0.4~~~& 99.6~~~& 99.6~~~& 100.0~~~\\
~1024~~~~~~& 13.8~~~& 96.5~~~& 96.8~~~& 100.0~~~\\
32768~~~~~~& 59.0~~~& 87.9~~~& 88.6~~~& 100.0~~~\\
\hline
\end{tabular}
\vspace{4mm}
\caption{Cumulative percentages of the execution times for the successive stages 
of \textsf{BbST\textsubscript{CON}} with the fastest serial sort (kxsort).
The default value of $k$ (512) was used.
Each row stands for a different number of queries (given in thousands).}
\label{table:partial}
\end{table}

\begin{figure}[pt]
\centerline{
\includegraphics[width=0.495\textwidth,scale=1.0]{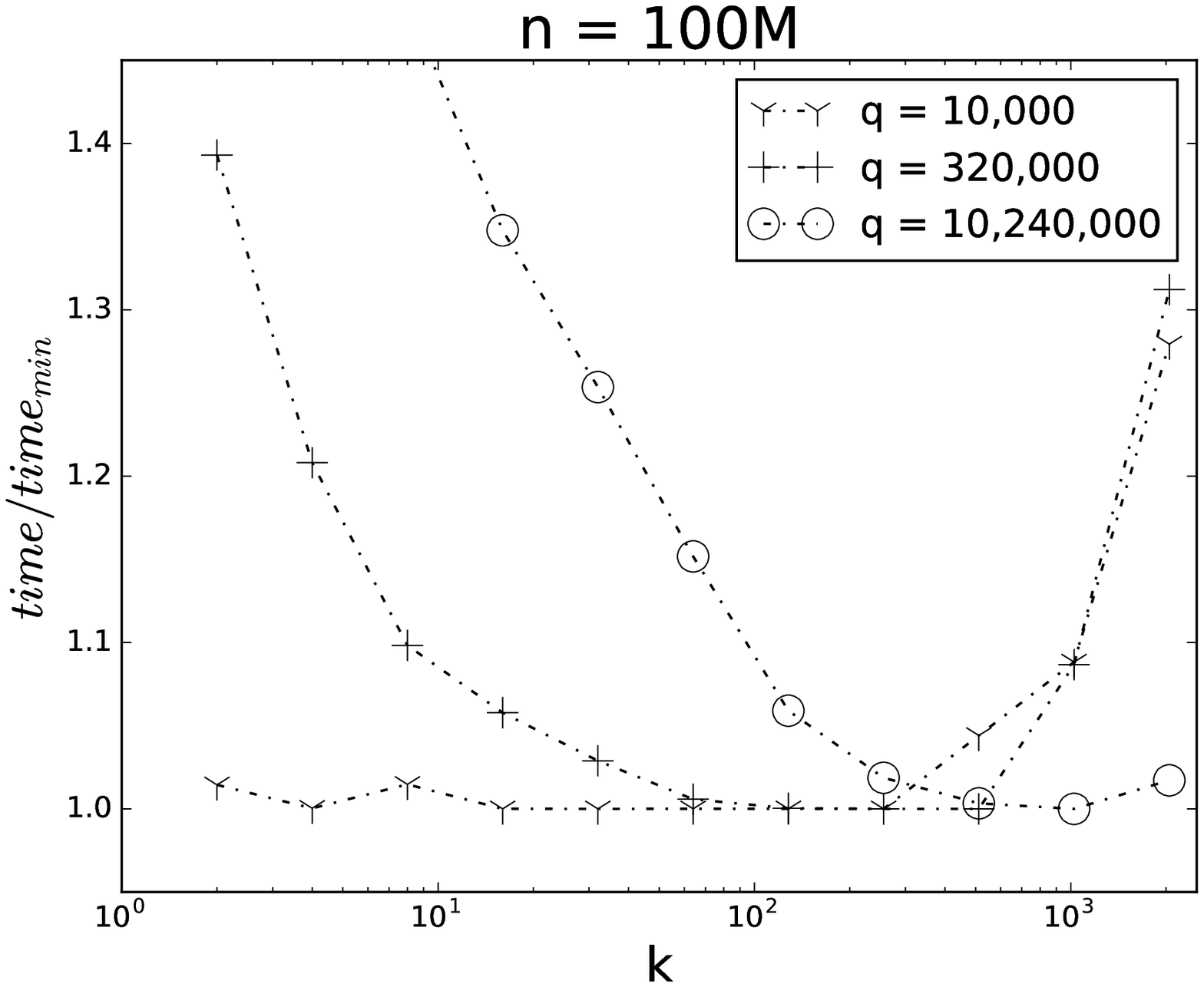}
\includegraphics[width=0.495\textwidth,scale=1.0]{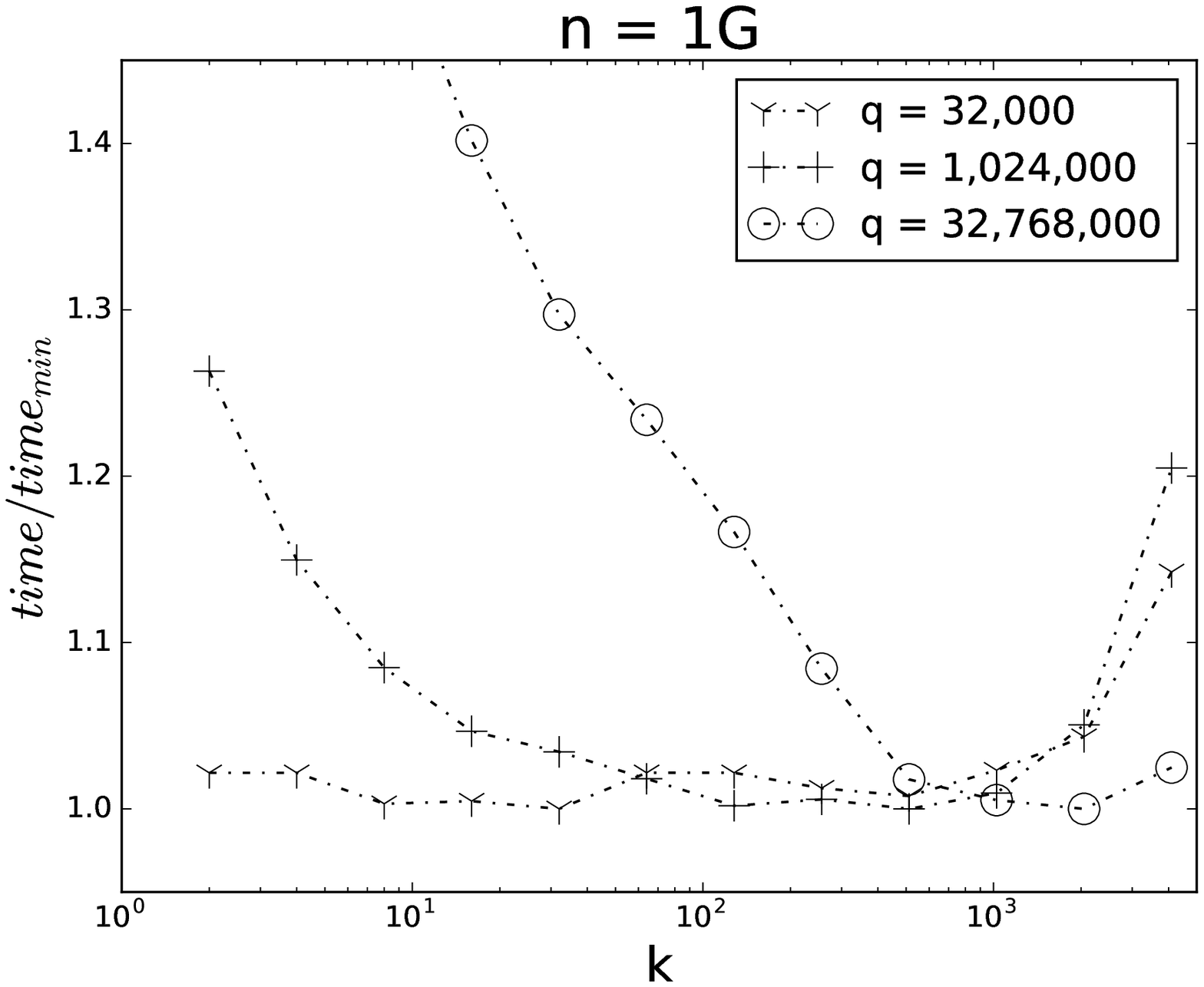}
}
\caption[Fig2]
{Ratio of running times to minimal running time for \textsf{BbST\textsubscript{CON}}
for several values of the block size $k$ and 
varying the number of queries $q$, from $\sqrt{n}$ to $1024\sqrt{n}$, 
where $n$ is 100 million (left figure) or 1 billion (right figure)}
\label{fig:var_k}
\end{figure}

In Fig.~\ref{fig:var_k} we varied the block size $k$ (the default sort, kxsort, 
was used).
With a small number of queries the overall timings are less sensitive to the 
choice of $k$. It is interesting to note that optimal $k$ can be found significantly below $\sqrt{n}$.

\begin{figure}[pt]
\centerline{
\includegraphics[width=0.495\textwidth,scale=1.0]{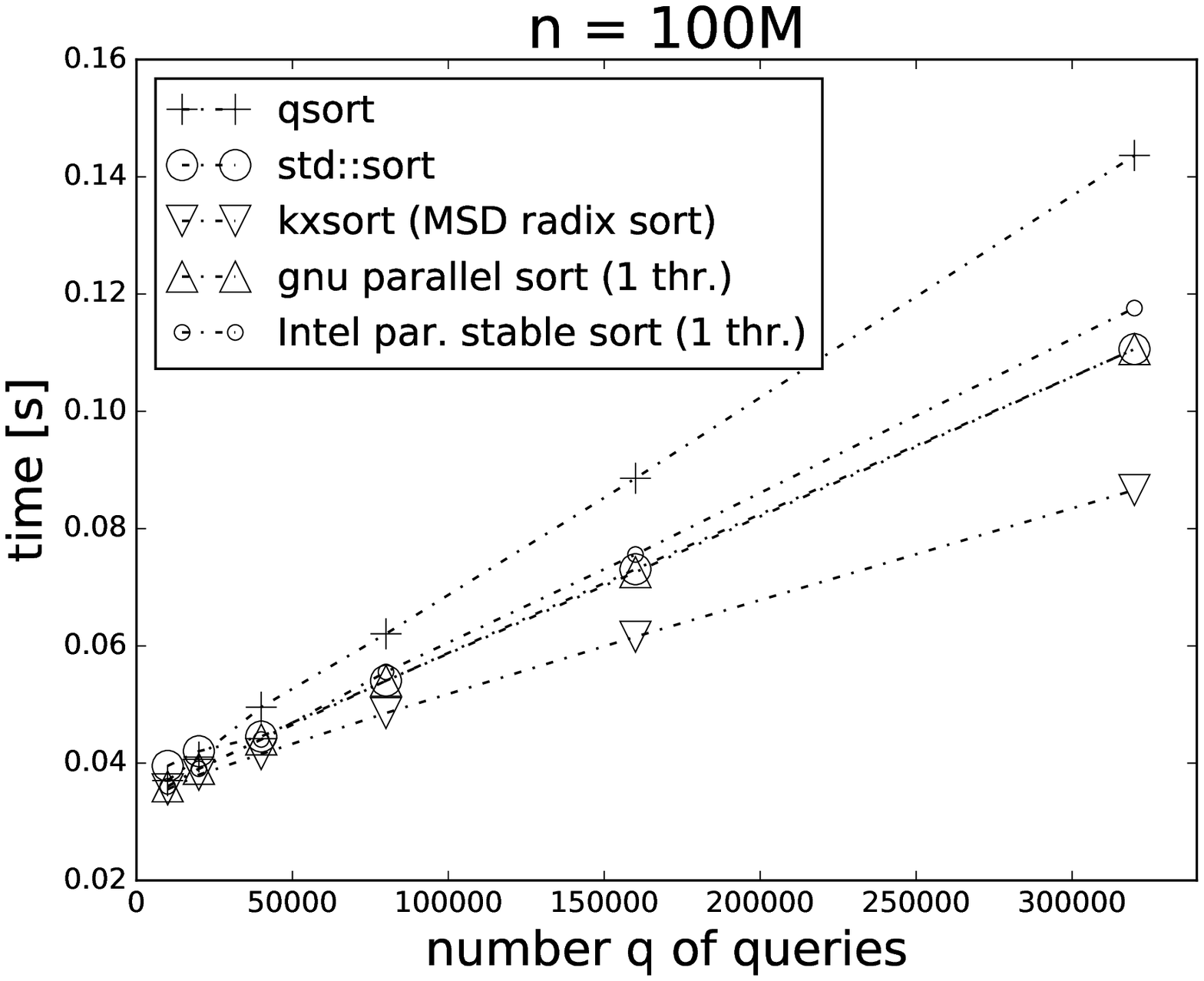}
\includegraphics[width=0.495\textwidth,scale=1.0]{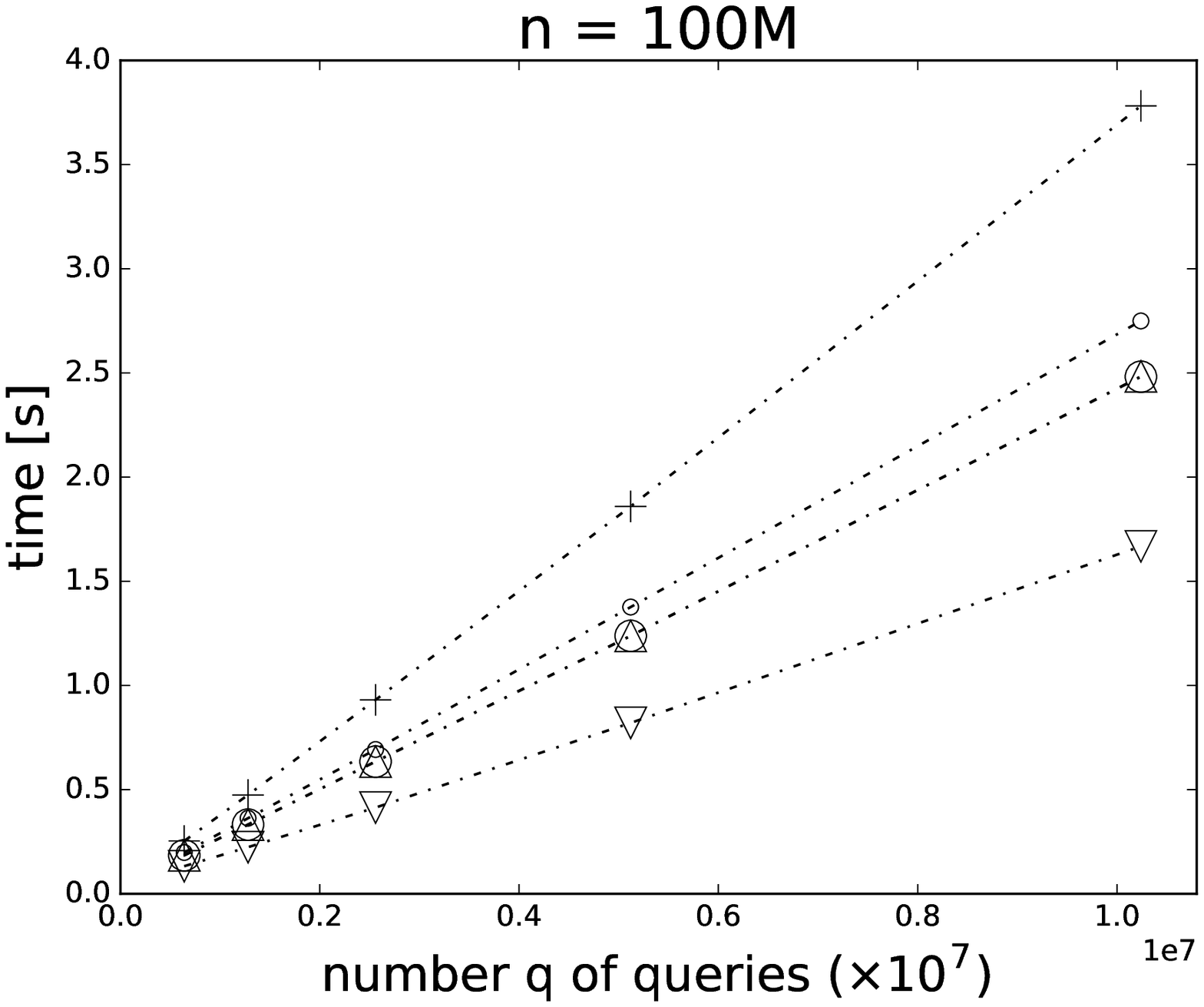}
}
\centerline{
\includegraphics[width=0.495\textwidth,scale=1.0]{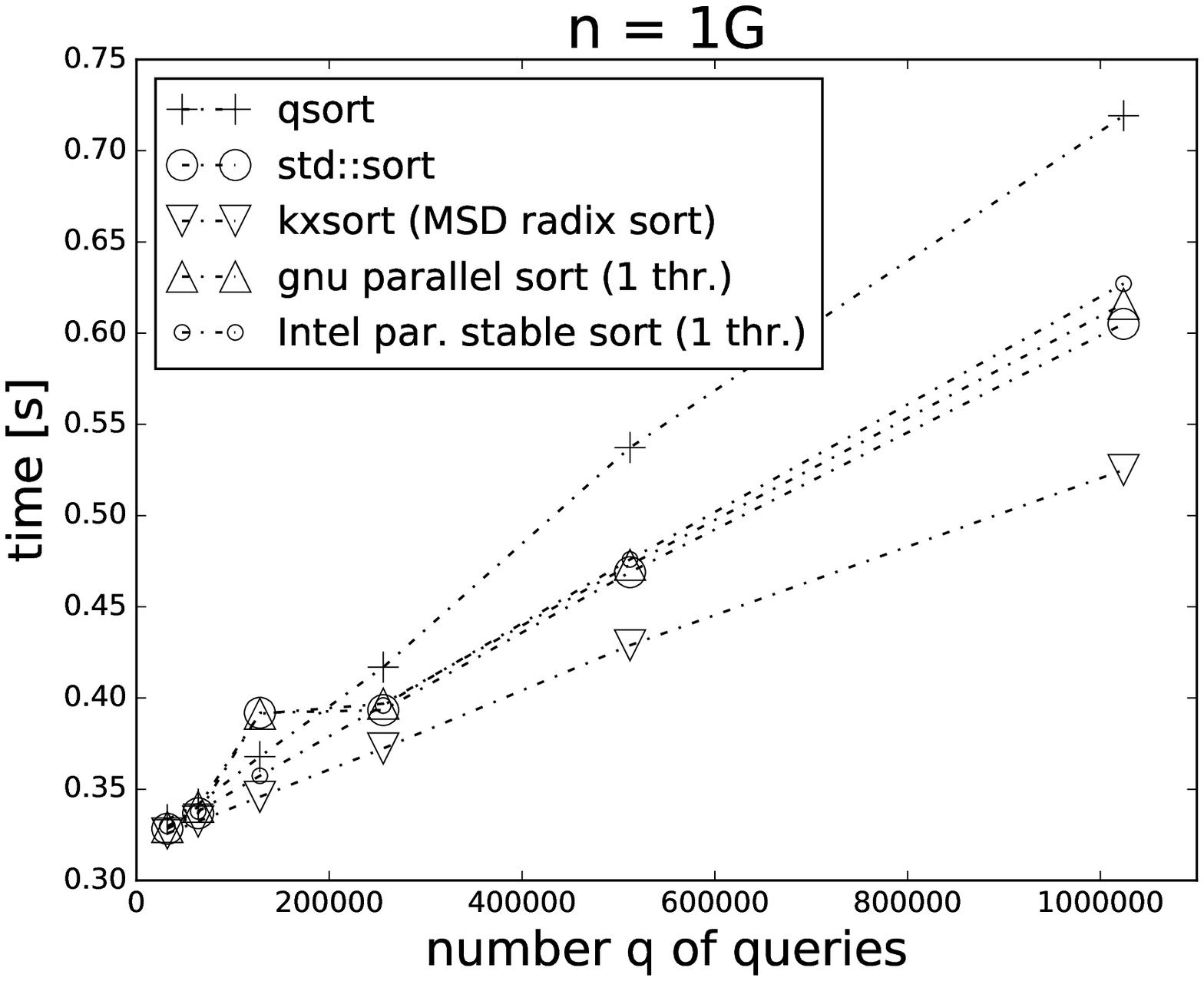}
\includegraphics[width=0.495\textwidth,scale=1.0]{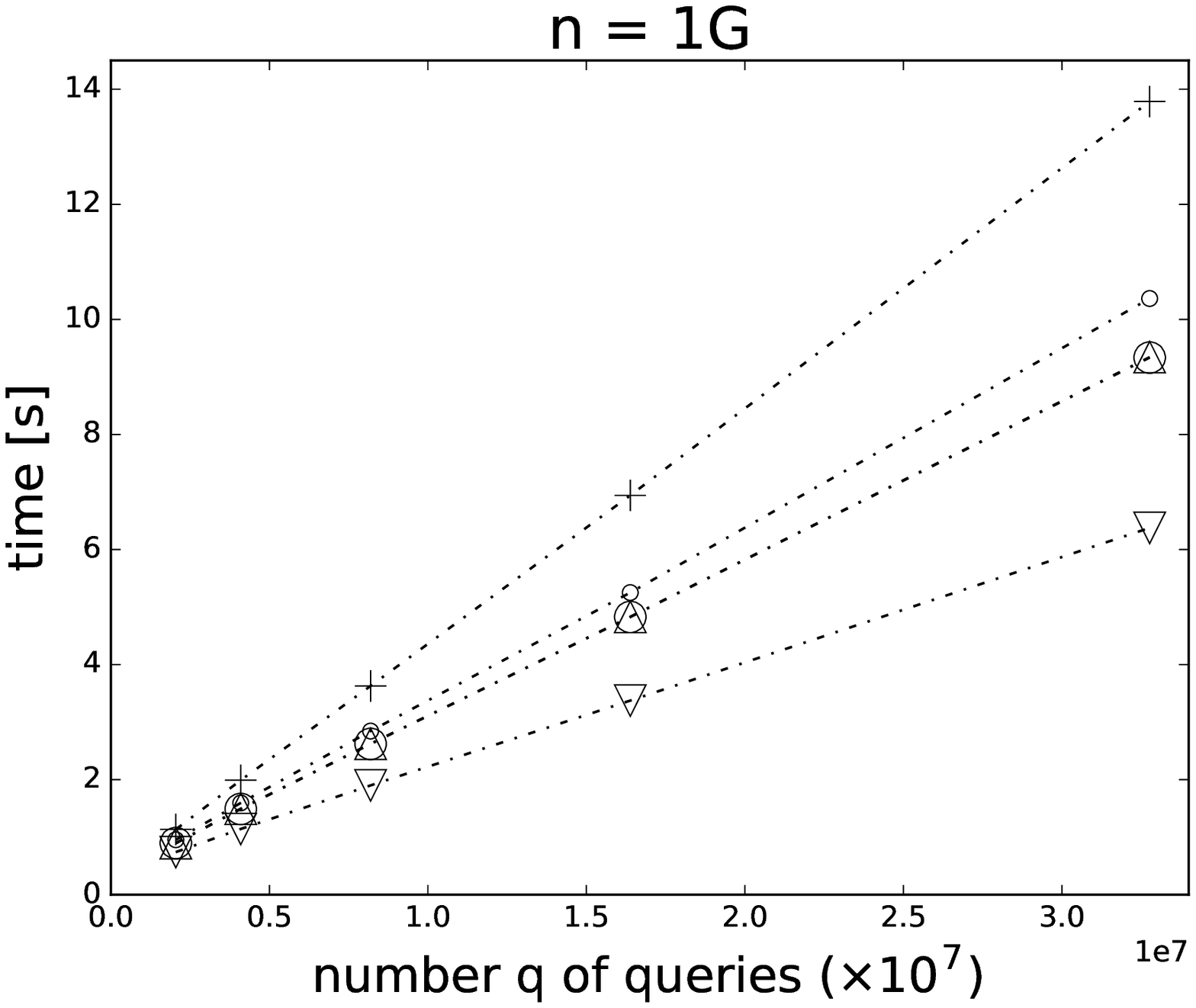}
}
\caption[Fig3]
{Impact of the sort algorithm on the running times of \textsf{BbST\textsubscript{CON}}.
The number of queries $q$ varies from $\sqrt{n}$ to $32\sqrt{n}$ (left figures) 
and from $64\sqrt{n}$ to $1024\sqrt{n}$ (right figures), 
where $n$ is 100 million (top figures) or 1 billion (bottom figures)}
\label{fig:var_sort}
\end{figure}

Different sorts, in a serial regime, were applied in the experiment shown in 
Fig.~\ref{fig:var_sort}.
Namely, we tried out C++'s qsort and std::sort, kxsort, \_\_gnu\_parallel::sort 
and Intel parallel stable sort (pss).
The function qsort, as it is easy to guess, is based on quick sort.
The other sort from the C++ standard library, std::sort, implements introsort, 
which is a hybrid of quick sort and heap sort.
Its idea is to run quick sort and only if it gets into trouble on some pathological data 
(which is detected when the recursion stack exceeds some threshold), switch to heap sort.
In this way, std::sort works in $O(n\log n)$ time in the worst case.
The next contender, kxsort, is an efficient MSD radix sort.
The last two sorters are parallel algorithms, but for this test they are run with 
a single thread.
The gnu sort is a multiway mergesort (exact variant)
from the GNU libstdc++ parallel mode library.
Finally, Intel's pss is a parallel merge 
sort\footnote{\texttt{https://software.intel.com/en-us/articles/\\
a-parallel-stable-sort-using-c11-for-tbb-cilk-plus-and-openmp}}.
We use it in the OpenMP~3.0 version.

\begin{figure}[pt!]
\centerline{
\includegraphics[width=0.495\textwidth,scale=1.0]{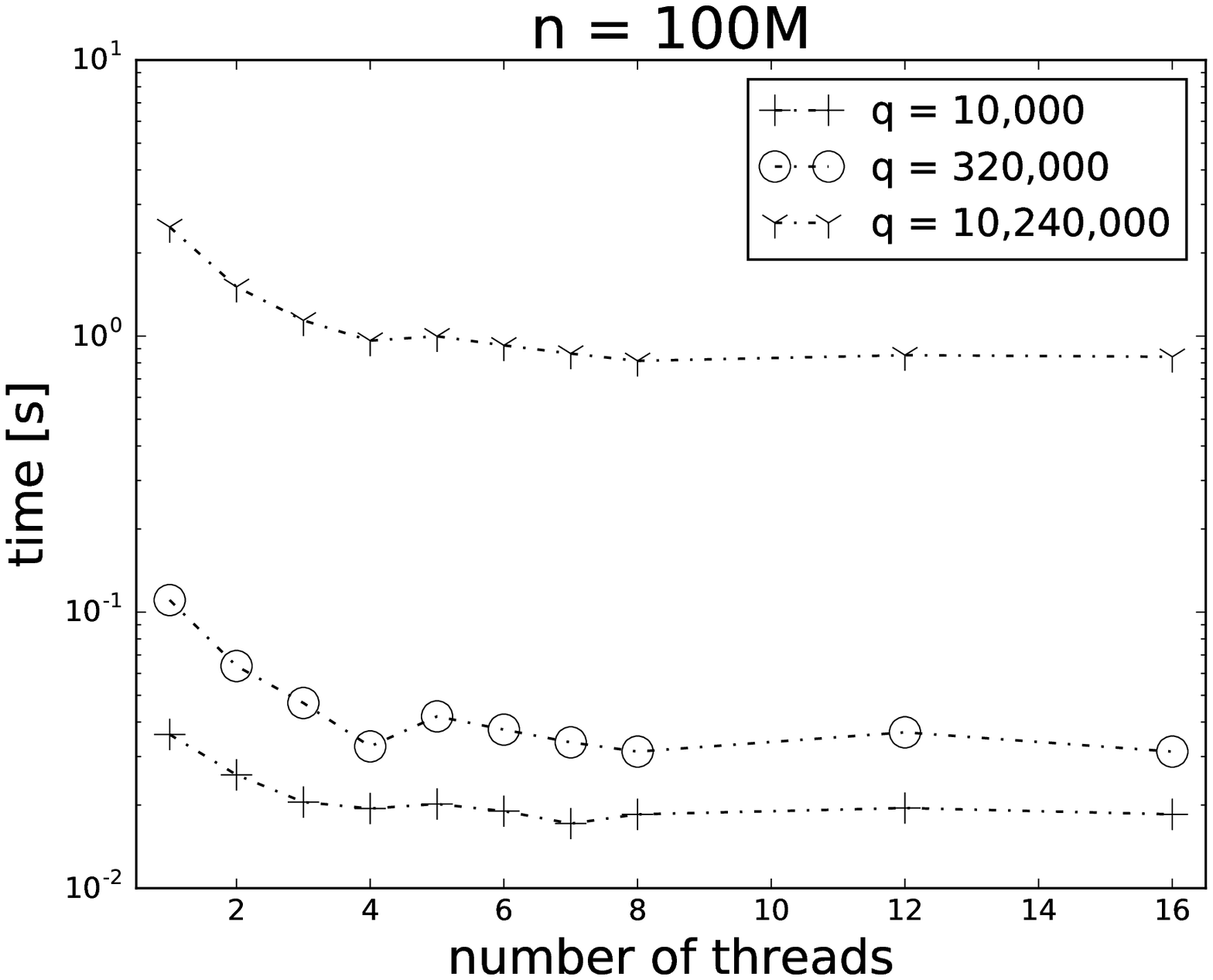}
\includegraphics[width=0.495\textwidth,scale=1.0]{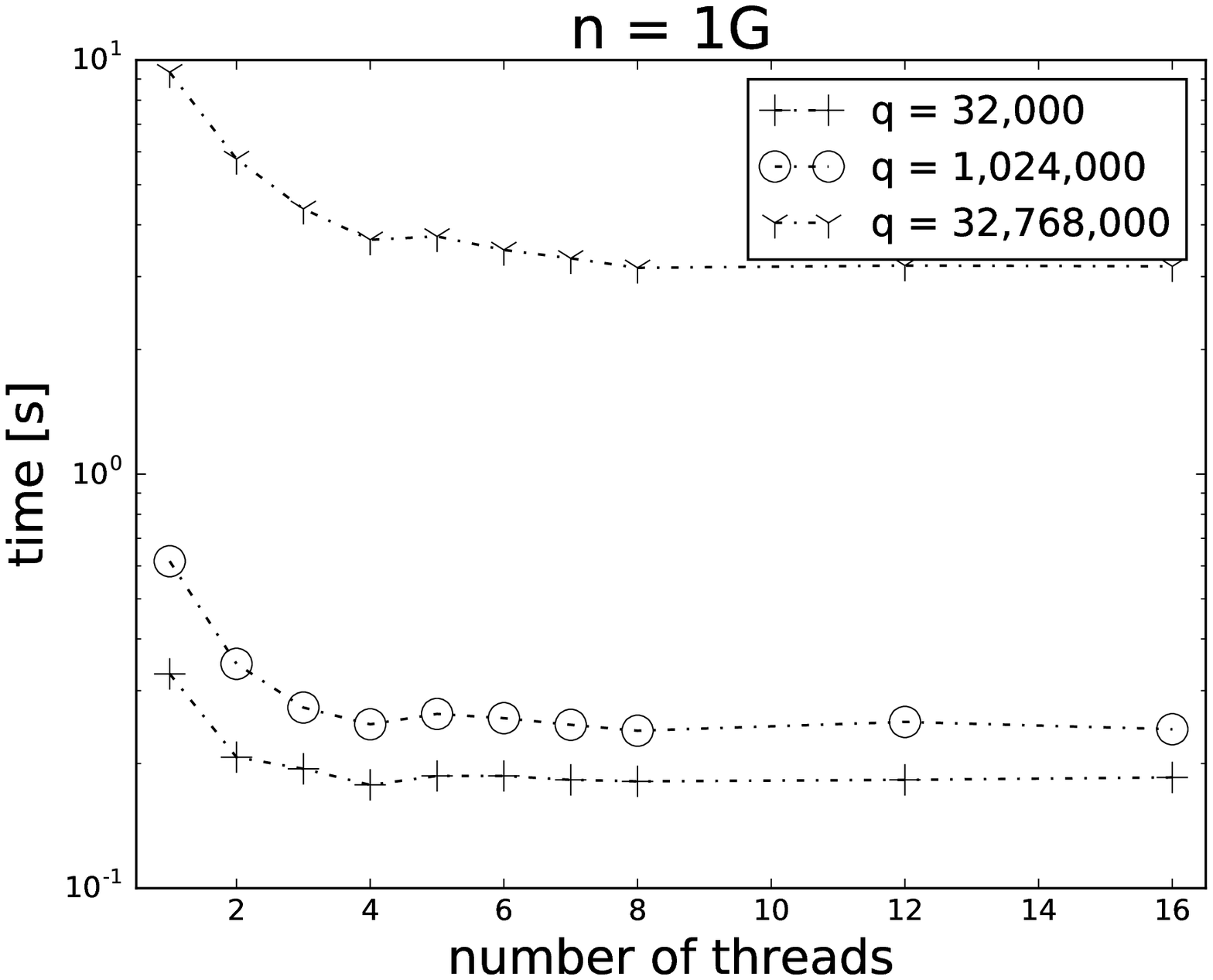}
}
\caption[Fig4]
{Impact of the number of threads in \_\_gnu\_parallel::sort 
and in creating $A_Q$ (by independent scanning for minima in 
contiguous areas of $A$)
on the overall performance of \textsf{BbST\textsubscript{CON}}, 
for different number of queries $q$, 
where $n$ is 100 million (left figure) or 1 billion (right figure).
Note the logarithmic scale on the Y-axis.}
\label{fig:var_threads}
\end{figure}

For the last experiment with \textsf{BbST\textsubscript{CON}}, we ran our algorithm in a parallel mode, 
varying the number of threads in $\{1, 2, \ldots, 8, 12, 16\}$ (Fig~\ref{fig:var_threads}).
For sorting the queries we took the faster parallel sort, 
\_\_gnu\_parallel::sort.
The remaining stages also benefit from parallelism.
The second stage computes in parallel the minima in contiguous areas of $A$ and the third stage correspondingly handles blocks of $A_Q$.
Finally, answering queries is handled in an embarassingly parallel manner.
As expected, the performance improves up to 8 threads 
(as the test machine has 4 cores and 8 hardware threads), but the overall 
speedups compared to the serial variant are rather disappointing, 
around factor 2 or slightly more.

\begin{figure}[pt!]
\centerline{
\includegraphics[width=0.495\textwidth,scale=1.0]{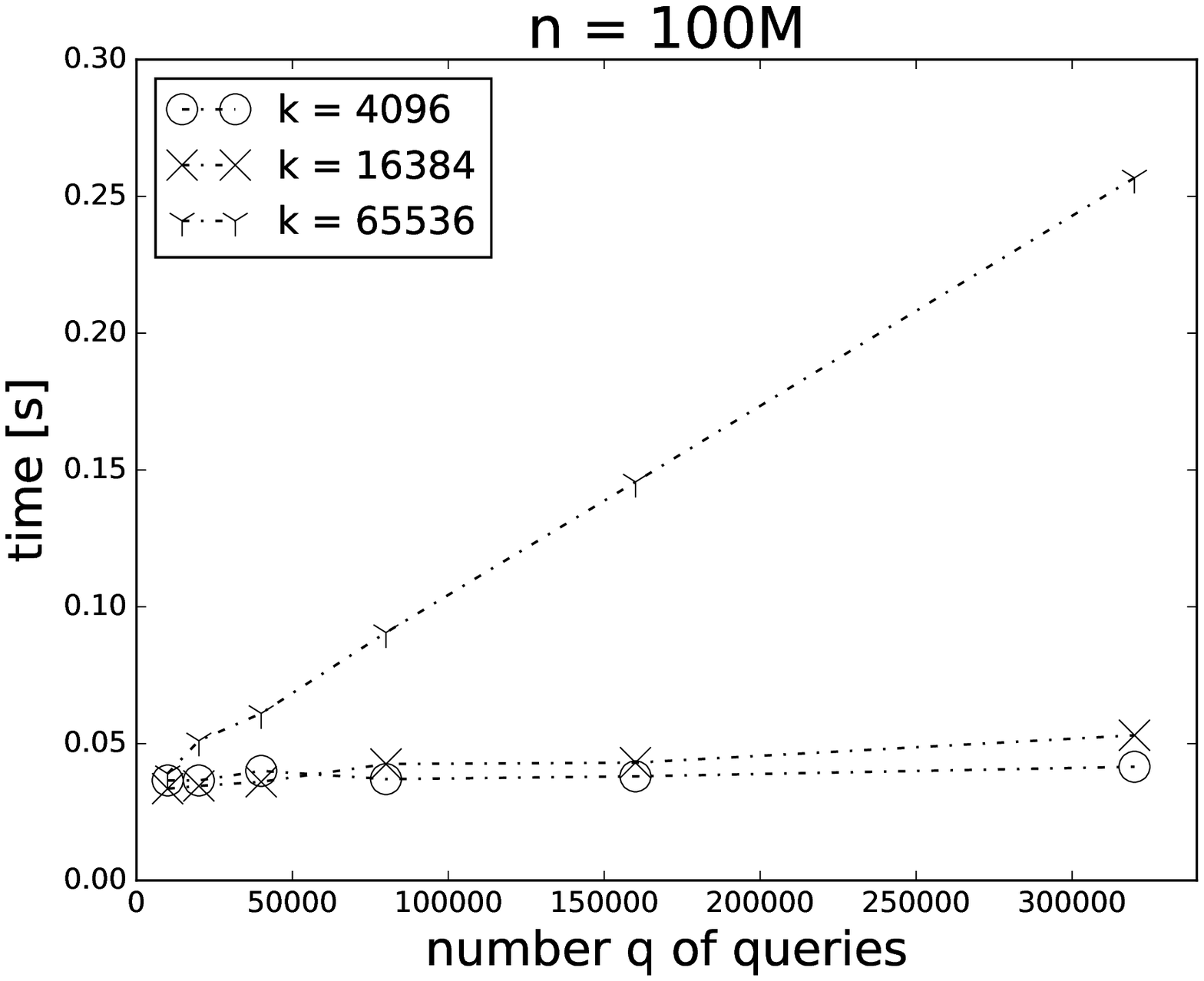}
\includegraphics[width=0.495\textwidth,scale=1.0]{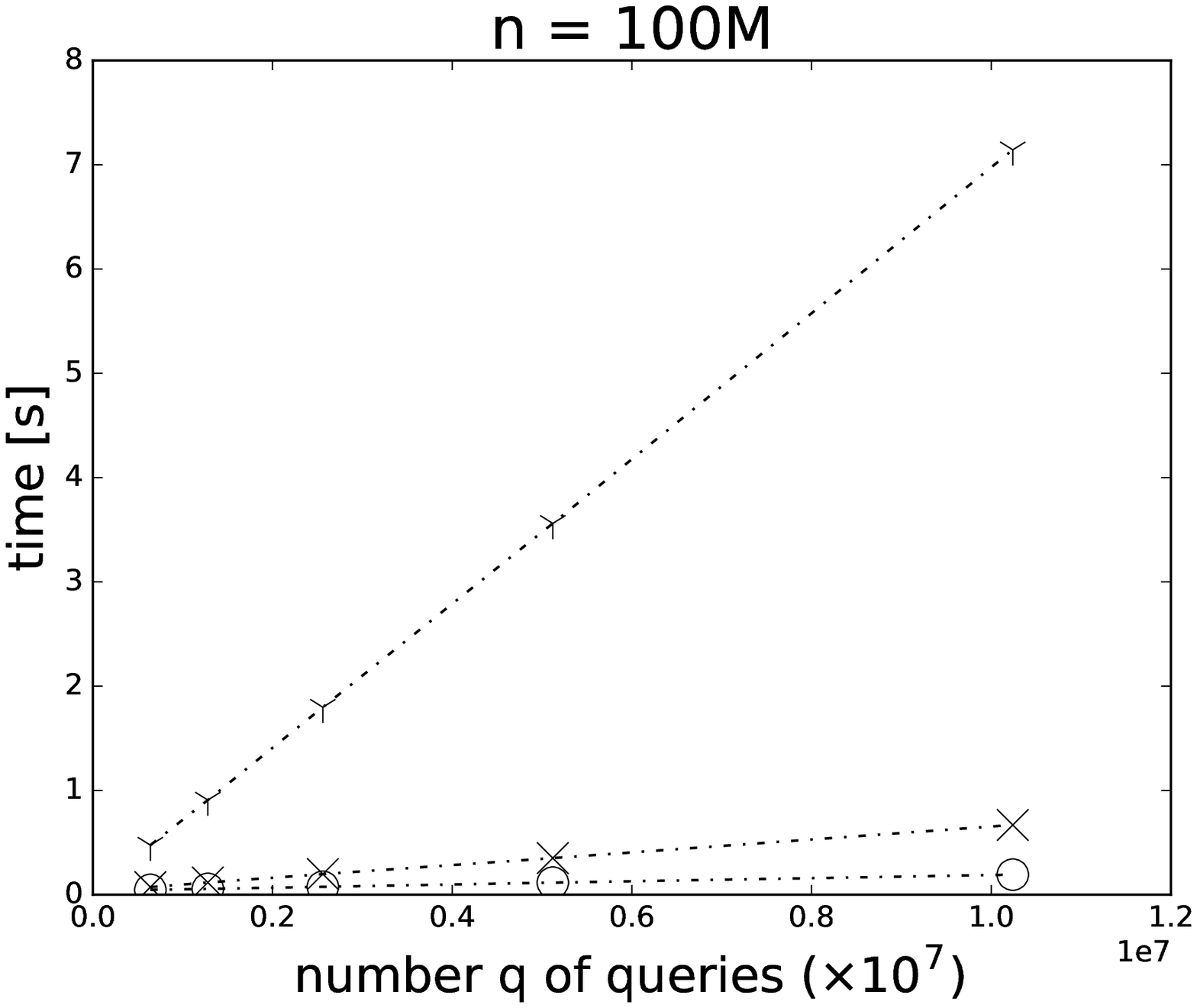}
}
\centerline{
\includegraphics[width=0.495\textwidth,scale=1.0]{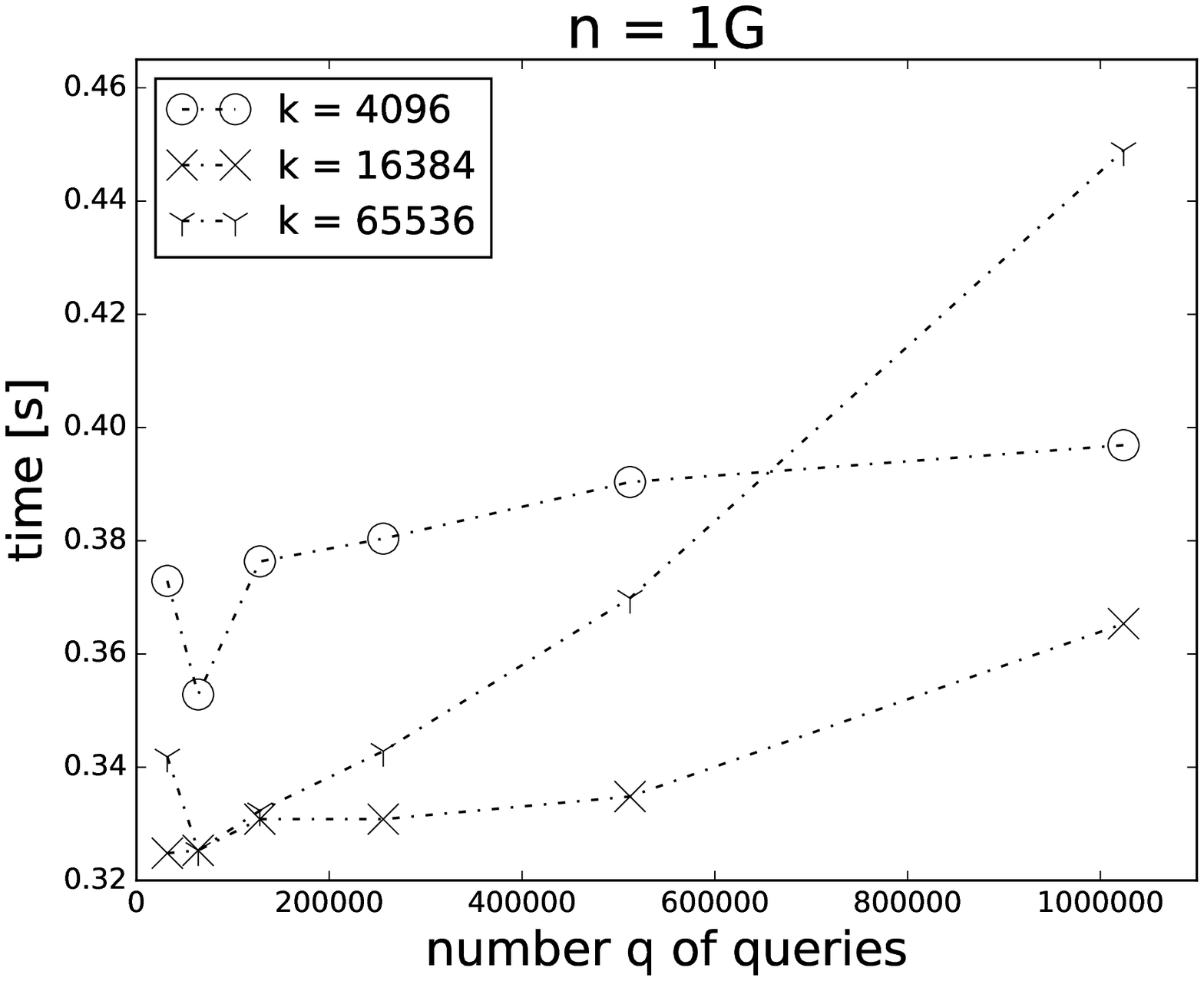}
\includegraphics[width=0.495\textwidth,scale=1.0]{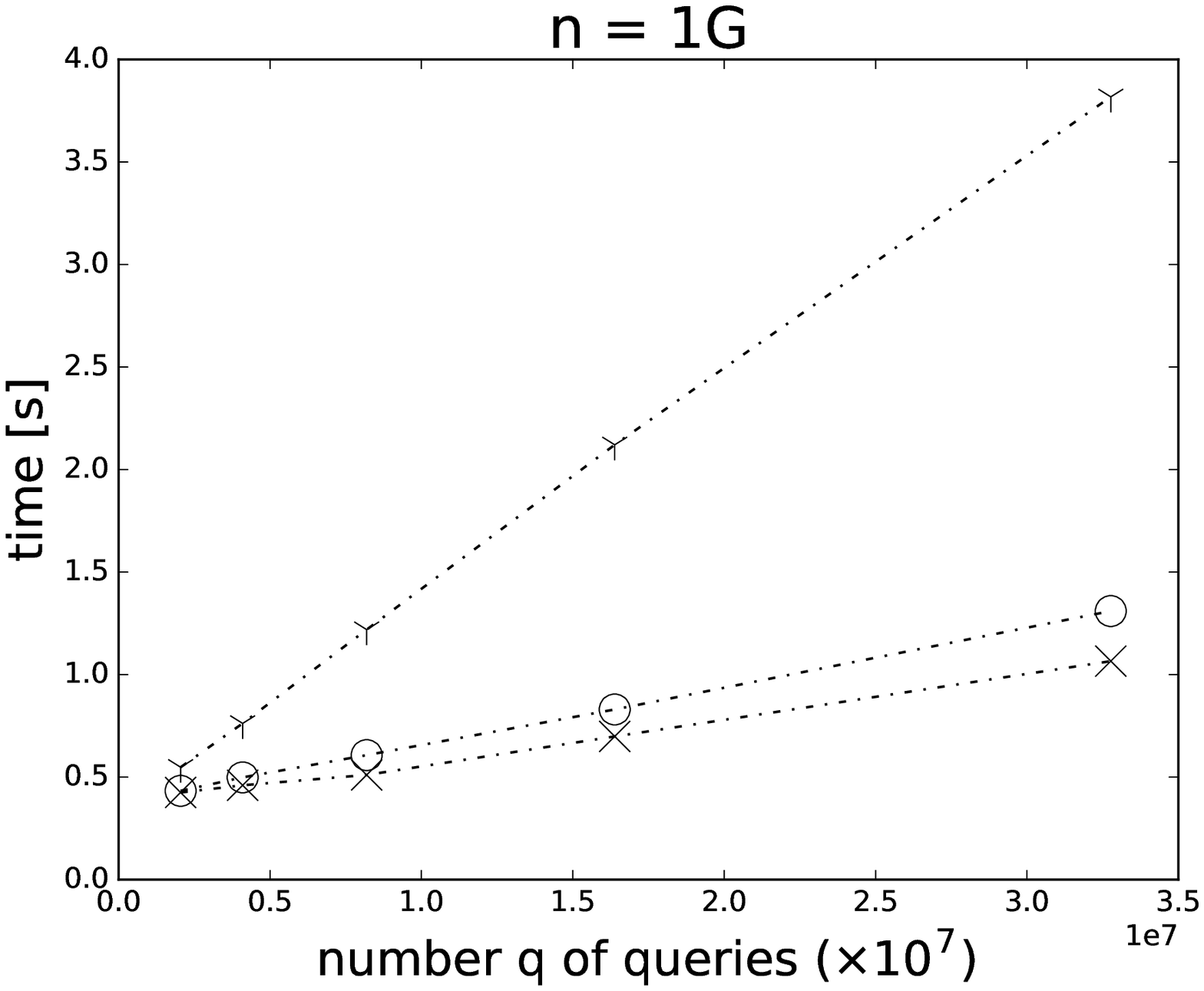}
}
\caption[Fig5]
{Running times for \textsf{BbST}
for several values of the block size $k$ and 
varying the number of queries $q$, 
from $\sqrt{n}$ to $32\sqrt{n}$ (left figures) 
and from $64\sqrt{n}$ to $1024\sqrt{n}$ (right figures), 
where $n$ is 100 million (top figures) or 1 billion (bottom figures)}
\label{fig:sb-rmq2}
\end{figure}

Finally, we ran a preliminary test of the algorithm from 
Subsection~\ref{sec:our2}, \textsf{BbST}, 
using the parameters of $k = \{4096, 16384, 65536\}$ (Fig.~\ref{fig:sb-rmq2}).
As expected, a smaller value of $k$ fits better the smaller value of $n$ 
and vice versa
(but for small $q$ and the larger $n$ our timings were slightly unpredictable).
Although we have not tried to fine tune the parameter $k$, we can easily 
see the potential of this algorithm.
For example, with $k = 16384$ and the largest tested number of queries, 
\textsf{BbST} is 2.5~times faster than \textsf{BbST\textsubscript{CON}} 
for the smaller $n$ and almost 6~times faster for the larger $n$.
Changing $k$ to $4096$ in the former case increases the time ratio 
to well over 8-fold!

Table~\ref{table:memory} presents the memory use 
(apart from input array $A$ and the set of queries $Q$) for the two variants.
\textsf{BbST} is insensitive here to $q$.
The parameter $k$ was set to 512 in the case of \textsf{BbST\textsubscript{CON}}. 
As expected, the space for \textsf{BbST\textsubscript{CON}} 
grows linearly with $q$.
\textsf{BbST} is more succinct for the tested number of queries 
($q \geq \sqrt{n}$), even if for a very small $q$ 
\textsf{BbST\textsubscript{CON}} would easily win in this respect.

\begin{table}[t!]
\centering
\begin{tabular}{lrr}
\hline
variant~~~~~ & \multicolumn{2}{c}{extra space as \% of the input} \\
with parameter &~~~~~$n = 100,000,000$ &~~~~~$n = 1,000,000,000$~\\
\hline
\textsf{BbST\textsubscript{CON}}, $q \approx \sqrt{n}$ & 0.10 & $ 0.03 $ \\
\textsf{BbST\textsubscript{CON}}, $q \approx 32 \sqrt{n}$ & 3.23 & 1.03 \\
\textsf{BbST\textsubscript{CON}}, $q \approx 1024 \sqrt{n}$ & 103.68 & 33.20 \\
\hline
\textsf{BbST}, $k = 2048$ & 1.56 & 1.86 \\
\textsf{BbST}, $k = 4096$ & 0.73 & 0.88 \\
\textsf{BbST}, $k = 8192$ & 0.34 & 0.42 \\
\textsf{BbST}, $k = 16,384$ & 0.16 & 0.20 \\
\textsf{BbST}, $k = 32,768$ & 0.07 & 0.09 \\
\hline
\end{tabular}
\vspace{4mm}
\caption{Memory use for the two variants, as the percentage of the 
space occupied by the input array $A$ (which is $4n$ bytes).
The parameter $k$ was set to 512 for \textsf{BbST\textsubscript{CON}}.
}
\label{table:memory}
\end{table}

\section{Final remarks}

We have proposed simple yet efficient algorithms for 
bulk range minimum queries.
Experiments on random permutations of $\{1, \ldots, n\}$ 
and with ranges chosen uniformly random over the input sequence 
show that one of our solutions, \textsf{BbST\textsubscript{CON}}, is from 3.8 to 7.8 times 
faster than its predecessor, \textsf{ST-RMQ\textsubscript{CON}} 
(the gap grows with increasing the number of queries).
The key idea that helped us achieve this advantage is adapting 
the well-known Sparse Table technique to work on blocks, 
with speculative block minima comparisons.

Not surprisingly, extra speedups can be obtained with parallelization, 
as shown by our preliminary experiments.
This line of research, however, should be pursued further.

The variant \textsf{BbST}, although possibly not as compact as \textsf{BbST\textsubscript{CON}} (when the number of queries is very small), 
proves even much faster.
We leave running more thorough experiments with this variant, 
including automated selection of parameter $k$,  
as a future work.

\section*{Acknowledgement}
The work was supported by the Polish National Science Centre under the project DEC-2013/09/B/ST6/03117 (both authors).

\bibliographystyle{psc}          
\bibliography{rmq}     

\end{document}